\documentclass[preprint,12pt]{elsarticle}
\usepackage{graphicx}
\usepackage{amssymb}
\usepackage{amsmath}
\usepackage{subfig}
\usepackage{multirow}
\usepackage{colortbl,color}
\usepackage{algorithm}
\usepackage[page,title,titletoc,header]{appendix}
\usepackage{multirow}
\usepackage{amsthm}
\usepackage{esint}
\usepackage{hyperref}
\usepackage{bm}
\biboptions{comma,sort&compress}
\journal{ }

\usepackage{cancel}
\begin{document}

\begin{frontmatter}
\title{Shear Stress and Fluid-Wall Interaction Force in LBM Simulations of Hydrodynamic and MHD Flows}
\author[a]{Jun~Li\footnote[1]{e-mail: lijun04@gmail.com}}
\address[a]{Institute of High Performance Computing (IHPC), \\ Agency for Science, Technology and Research (A*STAR), \\ 1 Fusionopolis Way, \#16-16 Connexis
North, Singapore 138632, Republic of Singapore}
\begin{abstract}
Chapman-Enskog analysis of the lattice Boltzmann method (LBM) is adopted to recover the Navier-Stokes (N-S) equation for the magnetohydrodynamic (MHD) flows driven by external body forces other than the induced Lorentz force. Various numerical schemes are discussed for the implementation of external body forces, leading to different artefact terms. An order-of-magnitude analysis is provided to demonstrate that the artefact terms are negligible in calculating the shear stress and strain rate tensors, which keeps the study of fluid-wall interaction and the implementation of large eddy simulation (LES) simple. This clarifies the confusion in the literature, where both considering and not considering the artefact terms have been adopted without discussion. Additionally, the fluid-wall interaction force can be computed as the intuitive momentum exchange rate using a special distribution function after a half-step propagation, and consistency is obtained with the formula derived using the Chapman-Enskog analysis. The momentum flux tensor for moving boundaries is uncovered as the cause of violating Galilean invariance and the Maxwell stress tensor for Lorentz force is also contained in this intuitively calculated interaction force for MHD flows due to the modification of equilibrium distribution function. Both extra terms should be removed for an interaction force associated with the hydrodynamic pressure and viscous shear stress. The implication of extra terms for the global balance of different physical forces is discussed.  
\end{abstract}
\end{frontmatter}

\section{Introduction}\label{Introduction}
We start the derivation and discussion with an introduction in Section~\ref{Basics} to the basic LBM algorithms used in flow simulations, which consider various force models, corresponding to different artefact terms in the recovered N-S equation. Confusion about the treatment of artefact terms in the stress tensor calculation is presented in Section~\ref{hydrodynamic stress tensor}. A detailed Chapman-Enskog analysis is provided in Section~\ref{MHD stress tensor}, showing that more complicated artefact terms could appear in MHD flow simulations, and an order-of-magnitude analysis is conducted to justify neglecting all artefact terms for numerical simplicity. A special distribution function after a half-step propagation is introduced in Section~\ref{interaction force}, which shows the consistency in computing the fluid-wall interaction force between an intuitive momentum exchange scheme and the formula obtained using Chapman-Enskog analysis. This clarifies why special corrections are required for moving boundaries and for some MHD models. Section~\ref{total force} presents the implication of the intuitively calculated interaction force for the global balance of different physical forces. Some conclusions are made in Section~\ref{Conclusions}.  

\section{Basic LBM algorithms}\label{Basics}

The computational domain is discretised using uniform grid points with a constant distance $\Delta x$ in all directions. The D2Q9 and D3Q19 lattice models \cite{Qianetal1992} are widely adopted in two-dimensional and three-dimensional flow simulations, respectively, to determine the lattice velocities $\vec e_\alpha$ and weight factors $\omega_\alpha$ for the $\alpha\in[0,Q-1]$ directions. The magnitude of $\vec e_\alpha$ depends on $c=\Delta x/\Delta t$ and $\Delta t$ is the timestep. Both $\Delta x$ and $\Delta t$ can take non-unity values in the current study. 

The only unknown is the density distribution function $f_\alpha$ and the corresponding equilibrium distribution function is defined as follows: 
\begin{equation} \label{eq:feq}
\begin{aligned}
   f^{\rm eq}_\alpha(\rho, \vec u^{\rm eq})=\rho\omega_\alpha\left[1+\dfrac{3}{c^2}\vec e_\alpha\cdot\vec u^{\rm eq}+\dfrac{9}{2c^4}(\vec e_\alpha\cdot\vec u^{\rm eq})^2-\dfrac{3}{2c^2}\vec u^{\rm eq}\cdot\vec u^{\rm eq}\right], 
\end{aligned}
\end{equation}
where the fluid density is computed as $\rho=\sum_\alpha f_\alpha$ but $\vec u^{\rm eq}$ depends on the adopted force models in the presence of external/internal body force $\vec a$ per unit mass, as introduced in the following. In particular,  $\vec u^{\rm eq}$ could be different from $\vec u=(1/\rho)\sum_\alpha\vec e_\alpha f_\alpha$ and the flow velocity $\vec v$ that can be computed as $\vec v=\vec u+0.5\Delta t\vec a$ for all force models.    

The distribution function $f_\alpha(\vec x,t)$ at the location $\vec x$ and time $t$ evolves according to an explicit relaxation-propagation algorithm for each $\Delta t$:
\begin{equation} \label{eq:evolution falpha}
\begin{split}
   f_\alpha(\vec x+\Delta t\vec e_\alpha,t+\Delta t)&=f_\alpha^{\rm rel}(\vec x,t) \\
   &=f_\alpha(\vec x,t)+\dfrac{f^{\rm eq}_\alpha(\rho, \vec u^{\rm eq})-f_\alpha(\vec x,t)}{\tau}+F_\alpha,
\end{split}
\end{equation}
where $f_\alpha^{\rm rel}$ is the distribution function after the relaxation and other local computations before propagating to the neighbouring grid points after $\Delta t$, the dimensionless relaxation time $\tau$ is selected according to the kinematic viscosity $\nu$ via $\nu=\Delta tc^2(\tau-0.5)/3$, and $F_\alpha$ is the force term used to implement any external or internal body forces. 

The simplest force model \cite{HeetalJSP1997} is $F_\alpha^{\rm [a]}=C\rho\omega_\alpha3\vec e_\alpha\cdot\vec a\Delta t/c^2$ with $C=1$ that has $\Theta_{ij}^{\rm [a]}=\sum_\alpha e_{\alpha,i}e_{\alpha,j}F_\alpha^{\rm [a]}=0$, where $i$ and $j$ are indices for the Cartesian coordinate directions. For body forces that change with time and spatial space, a refined force model is used to reduce discrete lattice effects \cite{MartysShanChenPRE1998, LuoPRE2000}: 
\begin{equation} \label{eq:force scheme}
\begin{split}
   & F_\alpha^{\rm [b]}=C\rho\omega_\alpha\left[\dfrac{3(\vec e_\alpha-\vec u^{\rm F})}{c^2}+\dfrac{9(\vec e_\alpha\cdot\vec u^{\rm F})\vec e_\alpha}{c^4}\right]\cdot\vec a\Delta t, \\
\end{split}
\end{equation}
which has: 
\begin{equation} \label{eq:force term requirement}
\begin{split}
   & \Theta_{ij}^{\rm [b]}=\sum_\alpha e_{\alpha,i}e_{\alpha,j}F_\alpha^{\rm [b]}=C\Delta t\rho(u_i^{\rm F}a_j+u_j^{\rm F}a_i), \\
\end{split}
\end{equation}
in addition to 
\begin{equation} \label{eq:force term requirement more}
\begin{split}
   & \sum_\alpha F_\alpha^{\rm [b]}=0, \\
   & \sum_\alpha e_{\alpha,i}F_\alpha^{\rm [b]}=C\Delta t\rho a_i.
\end{split}
\end{equation}
This model adopts $C=1$ and $\vec u^{\rm F}=\vec u$. The constraint of Eq.~\eqref{eq:force term requirement} is also satisfied in the integral form by the original force term of the Boltzmann equation \cite{LuoPRE2000}. Like in $F_\alpha^{\rm [b]}$, $\vec u$ is also used as the corresponding $\vec u^{\rm eq}$ of $f_\alpha^{\rm eq}$ in the relaxation term for momentum conservation since the force term already accounts for the total momentum increment due to the body force, i.e., $\sum_\alpha e_{\alpha,i} F_\alpha^{\rm [b]}=\Delta t\rho a_i$. 

The Shan-Chen force model \cite{ShanChenPRE1993} is originally constructed by having a shifted velocity $\vec u^{\rm eq}=\vec u+\tau\Delta t\vec a$ in $f_\alpha^{\rm eq}$ without the force term $F_\alpha$ but can be reformulated into the form of Eq.~\eqref{eq:evolution falpha} and the resulting force model $F_\alpha^{\rm [c]}$ \cite{LiLuoLiPRE2012} also satisfies Eqs.~\eqref{eq:force scheme}-\eqref{eq:force term requirement more}, having $C=1$, $\vec u^{\rm F}=\vec u+0.5\tau\Delta t\vec a$. After reformulation, $\vec u$ is used as the corresponding $\vec u^{\rm eq}$ of $f_\alpha^{\rm eq}$ in the relaxation term for momentum conservation.

Additionally, we also can implement the force using the exact difference method (EDM) \cite{Kupershtokh2009} and the original force term $F_\alpha^{\rm [d]}=f^{\rm eq}_\alpha(\rho, \vec u+\Delta t\vec a)-f^{\rm eq}_\alpha(\rho, \vec u)$ can be reformulated. After reformulation, the resulting force model $F_\alpha^{\rm [d]}$ \cite{LiLuoLiPRE2012} also satisfies Eqs.~\eqref{eq:force scheme}-\eqref{eq:force term requirement more}, having $C=1$, $\vec u^{\rm F}=\vec v$. Additionally, $\vec u$ is used as the corresponding $\vec u^{\rm eq}$ of $f_\alpha^{\rm eq}$ in the relaxation term for momentum conservation. The EDM and Shan-Chen force models with different $\vec u^{\rm F}$ become the same when $\tau=1$.  

To minimise discrete lattice effects, the artefact terms are eliminated for accurate recovery of the Navier-Stokes (N-S) equations up to the second-order time scale in the Chapman-Enskog analysis and the required force model $F_\alpha^{\rm [e]}$ \cite{GuoPRE2002} also satisfies Eqs.~\eqref{eq:force scheme}-\eqref{eq:force term requirement more}, having $C=[1-1/(2\tau)]\in(0,1)$, $\vec u^{\rm F}=\vec v$. Additionally, $\vec v$ is used as the corresponding $\vec u^{\rm eq}$ of $f_\alpha^{\rm eq}$ in the relaxation term.  

When all models are written in the form of Eq.~\eqref{eq:evolution falpha}, $C$ of $F_\alpha^{\rm [e]}$ is not unity since both the force and relaxation terms contribute to a total momentum increment $\Delta t\rho a_i$ after each $\Delta t$ while other force models with $C=1$ implement momentum increment only through the force term. Additionally, the flow velocity $\vec v$ is used in the force term $F_\alpha^{\rm [e]}$ as well as in the corresponding $f_\alpha^{\rm eq}$ for the relaxation term, which is also different from other force models \cite{GuoPRE2002, LiLuoLiPRE2012}. There are more force models that are not discussed in the current study, which is not intended to be exhaustive. Nevertheless, the following derivation is made general such that the result is also valid for other force models with given $F_\alpha$ and $\Theta_{ij}$. 

Although the momentum increment per unit volume after each $\Delta t$ by the force and relaxation terms is $\Delta t\rho a_i$, $\Theta_{ij}$ also contributes to the momentum increment since it is a momentum flux tensor, transporting $\Delta t\rho a_i$ at a velocity $u_j^{\rm F}$. Additionally, a similar momentum flux tensor with a transport velocity $u_j^{\rm eq}$ also appears in the recovered N-S momentum equation. These two extra contributions can be cancelled out when the force model is appropriately constructed but artefact terms will appear if not \cite{GuoPRE2002}. 

\section{Stress tensor calculations for hydrodynamic flows}\label{hydrodynamic stress tensor}
The Chapman-Enskog expansion can be used to recover the N-S equations from the LBM algorithms with different force models. For the force model $F_\alpha^{\rm [e]}$ \cite{GuoPRE2002}, the standard macroscopic equations are recovered as: 
\begin{equation} \label{eq:N-S mass for hydrodynamics}
\begin{split}
   & \dfrac{\partial \rho}{\partial t}+\nabla\cdot(\rho\vec u^{\rm eq})=0, \\
\end{split}
\end{equation}
and 
\begin{equation} \label{eq:N-S momentum for hydrodynamics}
\begin{split}
   & \dfrac{\partial (\rho\vec u^{\rm eq})}{\partial t}+\nabla\cdot(\rho\vec u^{\rm eq}\vec u^{\rm eq})=-\nabla p+\nabla\cdot(2\rho\nu\mathbf{S})+\rho\vec a, \\
\end{split}
\end{equation}
where $p=\rho c^2/3$ is pressure, the strain rate tensor $\mathbf{S}$ is recovered as $S_{ij}=(\partial u^{\rm eq}_j/\partial x_i+\partial u^{\rm eq}_i/\partial x_j)/2$ and $\vec u^{\rm eq}$ is equal to the flow velocity $\vec v$. 

For usual force models where $\vec u^{\rm eq}$ and $\vec u^{\rm F}$ could be different from $\vec v$, artefact terms will appear in the recovered macroscopic equations and numerical errors occur \cite{GuoPRE2002}, particularly in two-phase flow simulations \cite{LiLuoLiPRE2012}. However, once the force model for the evolution algorithm is fixed, the error of replacing $\vec v$ with $\vec u^{\rm eq}$ or $\vec u^{\rm F}$ in the incidental calculation of various quantities is negligible due to small $\Delta t$. For instance, we might replace the required $S_{ij}$ that is defined using $\vec v$ by the recovered $S_{ij}$ that depends on $\vec u^{\rm eq}$. We therefore neglect the possible but small error in determining $S_{ij}$ and compute the deviatoric/shear stress tensor using the same formula for all force models: 
\begin{equation} \label{eq:stress tensor for hydrodynamics}
\begin{split}
   & \sigma_{ij}=2\rho\nu S_{ij}\simeq\rho\nu(\dfrac{\partial u_j^{\rm eq}}{\partial x_i}+\dfrac{\partial u_i^{\rm eq}}{\partial x_j}), \\
\end{split}
\end{equation}
where $S_{ij}$ can be computed in LBM \cite{GuoPRE2002}: 
\begin{equation} \label{eq:strain rate tensor for hydrodynamics}
\begin{split}
   S_{ij}& =\dfrac{-3}{2\tau\Delta tc^2\rho}\left[\sum_\alpha e_{\alpha,i}e_{\alpha,j}(f_\alpha-f_\alpha^{\rm eq})+\tau\Delta t(u_i^{\rm eq}\rho a_j+u_j^{\rm eq}\rho a_i)-\tau\Theta_{ij}\right], \\
\end{split}
\end{equation}
where the term due to body force, $\tau\Delta t(u_i^{\rm eq}\rho a_j+u_j^{\rm eq}\rho a_i)-\tau\Theta_{ij}$, changes with the adopted force model and will be denoted by $E_{ij}$ for brevity. Although all force models can correctly implement the total momentum increment and therefore capture the dominant influence on flows, the corresponding $u_i^{\rm eq}$ and $\Theta_{ij}$ could be different. For incidental calculations of $S_{ij}$, we can save computational cost required for evaluating $f_\alpha^{\rm eq}$, namely replacing $\sum_\alpha e_{\alpha,i}e_{\alpha,j}(f_\alpha-f_\alpha^{\rm eq})$ with $\sum_\alpha e_{\alpha,i}e_{\alpha,j}f_\alpha-(p\delta_{ij}+\rho u_i^{\rm eq}u_j^{\rm eq})$. 

%\begin{equation} \label{eq:E_ij}
%E_{ij}= 
%\begin{cases}
%\tau\Delta t(u_i^{\rm eq}\rho a_j+u_j^{\rm eq}\rho a_i) & \text{if } F_\alpha=F_\alpha^{\rm [a]}, \\
%\dfrac{\Delta t}{2}(u_i^{\rm eq}\rho a_j+u_j^{\rm eq}\rho a_i) & \text{if } F_\alpha=F_\alpha^{\rm [e]}.
%\end{cases}
%\end{equation}

Note that the artefact term $E_{ij}$ contained in $\sigma_{ij}$ via $S_{ij}$ is neglected without clarification in Ref.~\cite{KrugerPRE2009} where the force model $F_\alpha^{\rm [a]}$ is adopted. In contrast, $E_{ij}$ of the force model $F_\alpha^{\rm [e]}$ is considered without discussion for simulations of non-Newtonian fluids in Ref.~\cite{WangAAMM2017} and for the large eddy simulation (LES) of hydrodynamic turbulence in Ref.~\cite{GuoJCP2025}. Additionally, the LES model is also applied to LBM simulations of electrohydrodynamic (EHD) turbulence in Ref.~\cite{ZhangPRF2024} but $E_{ij}$ of the force model $F_\alpha^{\rm [e]}$ is neglected without clarification. Moreover, the LES model can be used in LBM simulations of MHD turbulence and, as shown in the following derivation, more complicated artefact terms could appear in computing $\sigma_{ij}$, which is neglected without clarification in Refs.~\cite{DellarJCP2013, FlintVahala2017}. To clarify the confusion, the following derivation also provides an oder-of-magnitude analysis to justify neglecting all artefact terms in LBM simulations of MHD, EHD and hydrodynamic flows.

\section{Stress tensor calculations for MHD flows}\label{MHD stress tensor}
We extend the Chapman-Enskog analysis developed for hydrodynamic flows with a body force \cite{GuoPRE2002} to MHD flows. Note that the Chapman-Enskog analysis for MHD flows was given in \cite{Dellar2002} where the induced Lorentz force as well as a special body force is implemented by modifying $f_\alpha^{\rm eq}$ for the relaxation term and therefore the force term $F_\alpha$ is not considered. We consider the general approach that includes $F_\alpha$ for the external body force and  modification of $f_\alpha^{\rm eq}$ for the variable Lorentz force, as used in \cite{LietalFED2025} for fusion related applications, and the corresponding calculation of stress tensor is discussed in the current work. The third approach for MHD simulations is that both the Lorentz force and other body force are implemented using $F_\alpha$ and therefore $f_\alpha^{\rm eq}$ keeps unchanged, as adopted in \cite{PattisonFED2008}, for which the calculation of shear stress tensor reduces to Eqs.~\eqref{eq:stress tensor for hydrodynamics}-\eqref{eq:strain rate tensor for hydrodynamics}, as discussed in Section~\ref{hydrodynamic stress tensor}.  

For MHD flows driven by an external body force $\vec a$, the induced Lorentz force changes with time and the space coordinates. To reduce the artefact due to discrete lattice effects \cite{Dellar2002}, $f_\alpha^{\rm eq}$ in the relaxation term is modified to incorporate the Lorentz force as minus the divergence of Maxwell stress tensor. In this case, the evolution algorithm of Eq.~\eqref{eq:evolution falpha} is unchanged but $f^{\rm eq}_\alpha$ takes the following form for three-dimensional problems \cite{VahalaCCP2008, Dellar2009, LietalAMM2024}: 
 \begin{equation} \label{eq:feq for MHD}
\begin{aligned}
   f^{\rm eq}_\alpha(\rho, \vec u^{\rm eq}, \vec B)=&\rho\omega_\alpha\left[1+\dfrac{3}{c^2}\vec e_\alpha\cdot\vec u^{\rm eq}+\dfrac{9}{2c^4}(\vec e_\alpha\cdot\vec u^{\rm eq})^2-\dfrac{3}{2c^2}\vec u^{\rm eq}\cdot\vec u^{\rm eq}\right] \\
    &+\dfrac{9\omega_\alpha}{2c^4\mu}\left[\dfrac{B^2}{2}(\vec e_\alpha\cdot\vec e_\alpha)-(\vec e_\alpha\cdot\vec B)^2-\dfrac{c^2B^2}{6}\right], 
\end{aligned}
\end{equation}
where $\mu$ is the magnetic permeability of fluid. The modification of $f^{\rm eq}_\alpha$ for two-dimensional problems should exclude $-c^2B^2/6$ from Eq.~\eqref{eq:feq for MHD}, as detailed in the seminal work by Dellar \cite{Dellar2002}. The magnetic field $\vec B$ with a magnitude $B$ is two-way coupled with the flow simulation and solved using a separate vector-valued LBM algorithm for the magnetic induction equation, which is detailed in \cite{Dellar2002}. The current work is focused on the LBM algorithm for the flow simulation and the vector-valued LBM algorithm for $\vec B$ is omitted for brevity. Different LBM algorithms for simulating various physical problems are analogous and essentially work as solvers of a generalised governing equation \cite{LietalAMM2025}. 
 
As used to differentiate from $\vec u$, we have $\vec u^{\rm eq}$ in $f_\alpha^{\rm eq}$ for the relaxation term and $\vec u^{\rm F}$ in the force term $F_\alpha$, in addition to having a model-dependent coefficient $C$ in $F_\alpha$. $\vec u^{\rm F}$ is contained in $\Theta_{ij}$ in the recovered equations. 
 
By a slight abuse of notation for scalar, vector and tensors, we denote the $i$-th order moment of $f_\alpha^{\rm eq}$ by $M^{\rm eq}$ without or with different subscripts: 
\begin{equation} \label{eq:M of feq}
\begin{split}
   & M^{\rm eq}=\sum_\alpha f_\alpha^{\rm eq}=\rho, \\
   & M_i^{\rm eq}=\sum_\alpha e_{\alpha,i}f_\alpha^{\rm eq}=\rho u_i^{\rm eq}, \\
   & M_{ij}^{\rm eq}=\sum_\alpha e_{\alpha,i}e_{\alpha,j}f_\alpha^{\rm eq}=\dfrac{c^2}{3}\rho\delta_{ij}+\rho u_i^{\rm eq}u_j^{\rm eq}+L_{ij}, \\
   & M_{ijk}^{\rm eq}=\sum_\alpha e_{\alpha,i}e_{\alpha,j}e_{\alpha,k}f_\alpha^{\rm eq}=\dfrac{c^2}{3}\rho(\delta_{ij}u_k^{\rm eq}+\delta_{ik}u_j^{\rm eq}+\delta_{jk}u_i^{\rm eq}), \\
\end{split}
\end{equation}
where $\delta_{ij}$ is Kronecker delta and $L_{ij}=(1/\mu)(B^2\delta_{ij}/2-B_iB_j)$ is the Maxwell stress tensor corresponding to the Lorentz force. The modification to $f_\alpha^{\rm eq}$ by Eq.~\eqref{eq:feq for MHD} should only change $M_{ij}^{\rm eq}$, leading to an additional term $L_{ij}$.  

Note that the modification \cite{LiBrown2017} of $f^{\rm eq}_\alpha(\rho, \vec u^{\rm eq})$ to $\rho\omega_\alpha\left[1+3(\vec e_\alpha\cdot\vec u^{\rm eq})/c^2\right]$ only changes $M_{ij}^{\rm eq}$ to $c^2\rho\delta_{ij}/3$ and the correspondingly recovered equation is the Stokes equation without the convection term, which can be further extended to the Brinkman equation when a velocity-dependent body force is incorporated to account for the resistance of porous media \cite{GuoZhaoPRE2002, LiBrown2017}. Additionally, $f^{\rm eq}_\alpha(\rho, \vec u^{\rm eq})$ can be modified to change $M_{ij}^{\rm eq}$ and recover the desired bulk viscosity term in the N-S equation, as proposed in Ref.~\cite{Dellar2001}.

According to the Taylor expansion, we can rewrite Eq.~\eqref{eq:evolution falpha} into: 
\begin{equation} \label{eq:evolution falpha Taylor expansion}
\begin{split}
   & \sum_{n=1}^\infty\dfrac{\Delta t^n}{n!}D^n_tf_\alpha(\vec x,t)=\dfrac{f^{\rm eq}_\alpha(\rho, \vec u^{\rm eq})-f_\alpha(\vec x,t)}{\tau}+F_\alpha, \\
\end{split}
\end{equation}
where $D_t=(\partial_t+\vec e_\alpha\cdot\nabla)$. Now, the Chapman-Enskog expansion is applied: 
\begin{equation} \label{eq:Chapman-Enskog expansion}
\begin{split}
   & f_\alpha=f^{\rm eq}_\alpha+\sum_{n=1}^\infty f^{(n)}_\alpha, \\
   & \partial_t=\sum_{n=0}^\infty\partial_{t_n}. \\
\end{split}
\end{equation} 
Note that the expansion $\partial_t=\sum_{n=0}^\infty\partial_{t_n}$ of the time derivative is just a formal definition but not executable for any given analytical formula of $f_\alpha(\vec x,t)$ and thus this expansion has no tangible mathematical sense. 

Then, terms in Eq.~\eqref{eq:evolution falpha Taylor expansion} can be sorted according to the order of magnitude and Eq.~\eqref{eq:evolution falpha Taylor expansion} can be replaced by a series of equations arranged into a consecutive order of magnitude: 
\begin{equation} \label{eq:splited eqs}
\begin{split}
   & \Delta t(\partial_{t_0}+\vec e_\alpha\cdot\nabla)f^{\rm eq}_\alpha=\dfrac{-1}{\tau}f^{(1)}_\alpha+F_\alpha, \\
   & \Delta t(\partial_{t_0}+\vec e_\alpha\cdot\nabla)\left[(1-\dfrac{1}{2\tau})f^{(1)}_\alpha+\dfrac{F_\alpha}{2}\right]+\Delta t\partial_{t_1}f^{\rm eq}_\alpha=\dfrac{-1}{\tau}f^{(2)}_\alpha, \\
   & \cdots
\end{split}
\end{equation} 
where $F_\alpha$ without expansion is placed in the first equation as a dominant term. The original term, $0.5\Delta t^2(\partial_{t_0}+\vec e_\alpha\cdot\nabla)^2f^{\rm eq}_\alpha$, of the second equation has been replaced using the first one, leading to the magic coefficient $[1-1/(2\tau)]$. 

Considering $\sum_\alpha f_\alpha=\sum_\alpha f_\alpha^{\rm eq}=\rho$, $\sum_\alpha \vec e_\alpha f_\alpha=\rho\vec u$ and $\sum_\alpha \vec e_\alpha f_\alpha^{\rm eq}=\rho\vec u^{\rm eq}$, the following \textit{harsh} assumptions are adopted to make $f^{(n)}_\alpha$ of each $n$ tractable in Eq.~\eqref{eq:splited eqs}: 
\begin{equation} \label{eq:harsh conservation laws}
\begin{split}
   & \sum_\alpha f^{(n)}_\alpha=0, \quad \forall n\geq1, \\
   & \sum_\alpha e_{\alpha,i}f^{(1)}_\alpha=\rho(u_i-u_i^{\rm eq}), \\
   & \sum_\alpha e_{\alpha,i}f^{(n)}_\alpha=0, \quad \forall n\geq2. \\
\end{split}
\end{equation}

Using Eqs.~\eqref{eq:M of feq} and \eqref{eq:harsh conservation laws}, the zeroth-order moments of Eq.~\eqref{eq:splited eqs} are:  
\begin{equation} \label{eq:zero order NS}
\begin{split}
   & \dfrac{\partial\rho}{\partial t_0}+\dfrac{\partial(\rho u_j^{\rm eq})}{\partial x_j}=0, \\
   & (1-\dfrac{1}{2\tau})\dfrac{\partial(\rho u_j-\rho u_j^{\rm eq})}{\partial x_j}+C\dfrac{\partial(\rho\Delta t a_j/2)}{\partial x_j}+\dfrac{\partial\rho}{\partial t_1}=0. \\
\end{split}
\end{equation}
We consider possible improvement by always using $\vec v$ as the flow velocity for all force models, including the force model $F_\alpha^{\rm [e]}$ \cite{GuoPRE2002} with $C=[1-1/(2\tau)]$ and $\vec u^{\rm eq}=\vec v$ as well as other force models with $C=1$ and $\vec u^{\rm eq}=\vec u$. Then, the combination of the two equations of Eq.~\eqref{eq:zero order NS} using $\partial_{t_0}+\partial_{t_1}\approx\partial_t$ is the standard mass conservation equation:
\begin{equation} \label{eq:zero order NS-clean}
\begin{split}
   & \dfrac{\partial\rho}{\partial t}+\dfrac{\partial(\rho v_j)}{\partial x_j}=0. \\
\end{split}
\end{equation} 

For the recovery of the momentum equation, we can similarly compute the first-order moment of the first equation of Eq.~\eqref{eq:splited eqs}: 
\begin{equation} \label{eq:first order NS-part 1}
\begin{split}
   & \dfrac{\partial(\rho u_i^{\rm eq})}{\partial t_0}+\dfrac{\partial M_{ij}^{\rm eq}}{\partial x_j}=\rho a_i, \\
\end{split}
\end{equation}
where the original $\rho(u_i^{\rm eq}-u_i)/(\tau\Delta t)+C\rho a_i$ is replaced by $\rho a_i$, which should be satisfied by all force models for the total momentum increment contributed by the relaxation and force terms. Then, the first-order moment of the second equation of Eq.~\eqref{eq:splited eqs} is computed: 
\begin{equation} \label{eq:first order NS-part 2}
\begin{split}
   & (1-\dfrac{1}{2\tau})\left[\dfrac{\partial(\rho u_i-\rho u_i^{\rm eq})}{\partial t_0}+\dfrac{\partial}{\partial x_j}\sum_\alpha e_{\alpha,i}e_{\alpha,j}f^{(1)}_\alpha\right] \\
   & +C\dfrac{\partial(\rho\Delta t a_i/2)}{\partial t_0}+\dfrac{1}{2}\dfrac{\partial\Theta_{ij}}{\partial x_j}+\dfrac{\partial(\rho u_i^{\rm eq})}{\partial t_1}=0. \\
\end{split}
\end{equation}

For completeness, we continue to provide the details for the derivation such that the appearance of possible artefact terms is clear. Using Eqs.~\eqref{eq:M of feq}, \eqref{eq:splited eqs}, \eqref{eq:zero order NS} and \eqref{eq:first order NS-part 1}, we have: 
\begin{equation} \label{eq:long a}
\begin{split}
   & \sum_\alpha e_{\alpha,i}e_{\alpha,j}f^{(1)}_\alpha=-\tau\Delta t\sum_\alpha e_{\alpha,i}e_{\alpha,j}(\partial_{t_0}+\vec e_\alpha\cdot\nabla)f^{\rm eq}_\alpha+\tau\Theta_{ij} \\
   & =-\tau\Delta t\left[\dfrac{\partial}{\partial t_0}(\dfrac{c^2}{3}\rho\delta_{ij}+\rho u_i^{\rm eq}u_j^{\rm eq}+L_{ij})+\dfrac{\partial M_{ijk}^{\rm eq}}{\partial x_k}\right]+\tau\Theta_{ij} \\
   & =-\tau\Delta t\left[\cancel{\dfrac{-c^2}{3}\delta_{ij}\dfrac{\partial(\rho u_k^{\rm eq})}{\partial x_k}}+\dfrac{\partial(\rho u_i^{\rm eq}u_j^{\rm eq})}{\partial t_0}+\dfrac{\partial L_{ij}}{\partial t_0}+\dfrac{\partial M_{ijk}^{\rm eq}}{\partial x_k}\right]+\tau\Theta_{ij}, \\
\end{split}
\end{equation}
where the diagonal strikethrough line indicates cancellation of terms in Eqs.~\eqref{eq:long a}, \eqref{eq:long b} and \eqref{eq:long c} and
\begin{equation} \label{eq:long b}
\begin{split}
    \dfrac{\partial(\rho u_i^{\rm eq}u_j^{\rm eq})}{\partial t_0}=&u_i^{\rm eq}\dfrac{\partial(\rho u_j^{\rm eq})}{\partial t_0}+u_j^{\rm eq}\dfrac{\partial(\rho u_i^{\rm eq})}{\partial t_0}-u_i^{\rm eq}u_j^{\rm eq}\dfrac{\partial\rho}{\partial t_0} \\
    =&-u_i^{\rm eq}\dfrac{\partial M_{jk}^{\rm eq}}{\partial x_k}+u_i^{\rm eq}\rho a_j-u_j^{\rm eq}\dfrac{\partial M_{ik}^{\rm eq}}{\partial x_k}+u_j^{\rm eq}\rho a_i+u_i^{\rm eq}u_j^{\rm eq}\dfrac{\partial(\rho u_k^{\rm eq})}{\partial x_k} \\
    =&\cancel{-u_i^{\rm eq}\dfrac{c^2}{3}\dfrac{\partial\rho}{\partial x_j}}-\cancel{u_j^{\rm eq}\dfrac{c^2}{3}\dfrac{\partial\rho}{\partial x_i}}-\dfrac{\partial(\rho u_i^{\rm eq}u_j^{\rm eq}u_k^{\rm eq})}{\partial x_k}-u_i^{\rm eq}\dfrac{\partial L_{jk}}{\partial x_k}-u_j^{\rm eq}\dfrac{\partial L_{ik}}{\partial x_k} \\
   &+u_i^{\rm eq}\rho a_j+u_j^{\rm eq}\rho a_i
\end{split}
\end{equation}
and 
\begin{equation} \label{eq:long c}
\begin{split}
   & \dfrac{\partial M_{ijk}^{\rm eq}}{\partial x_k}=\dfrac{\partial}{\partial x_k}\left[\dfrac{c^2}{3}\rho(\delta_{ij}u_k^{\rm eq}+\delta_{ik}u_j^{\rm eq}+\delta_{jk}u_i^{\rm eq})\right] \\
   & =\cancel{\dfrac{c^2}{3}\delta_{ij}\dfrac{\partial(\rho u_k^{\rm eq})}{\partial x_k}}+\cancel{\dfrac{c^2}{3}\dfrac{\partial\rho}{\partial x_i}u_j^{\rm eq}}+\dfrac{c^2}{3}\rho\dfrac{\partial u_j^{\rm eq}}{\partial x_i}+\cancel{\dfrac{c^2}{3}\dfrac{\partial\rho}{\partial x_j}u_i^{\rm eq}}+\dfrac{c^2}{3}\rho\dfrac{\partial u_i^{\rm eq}}{\partial x_j}.
\end{split}
\end{equation}
Substituting Eqs.~\eqref{eq:long c} and \eqref{eq:long b} into Eq.~\eqref{eq:long a}, we obtain: 
\begin{equation} \label{eq:second-order moment of fneq}
\begin{split}
   \sum_\alpha e_{\alpha,i}e_{\alpha,j}f^{(1)}_\alpha=&-\tau\Delta t\left[\dfrac{c^2}{3}\rho(\dfrac{\partial u_j^{\rm eq}}{\partial x_i}+\dfrac{\partial u_i^{\rm eq}}{\partial x_j})-\dfrac{\partial(\rho u_i^{\rm eq}u_j^{\rm eq}u_k^{\rm eq})}{\partial x_k}\right] \\
   &-\tau\Delta t\left[-u_i^{\rm eq}\dfrac{\partial L_{jk}}{\partial x_k}-u_j^{\rm eq}\dfrac{\partial L_{ik}}{\partial x_k}+\dfrac{\partial L_{ij}}{\partial t_0}\right] \\
   &-\tau\Delta t(u_i^{\rm eq}\rho a_j+u_j^{\rm eq}\rho a_i)+\tau\Theta_{ij}.
\end{split}
\end{equation}

Before combining Eqs.~\eqref{eq:first order NS-part 1} and \eqref{eq:first order NS-part 2} for the momentum equation, we provide an order-of-magnitude analysis for possible simplifications to Eq.~\eqref{eq:second-order moment of fneq}: 
\begin{itemize}
    \item The higher-order term that occurs to all force models, $\partial(\rho u_i^{\rm eq}u_j^{\rm eq}u_k^{\rm eq})/\partial x_k$, is small at low Mach numbers (i.e., $|u_i^{\rm eq}|\ll c/\sqrt{3}$) and can be neglected, compared to other terms that are left in the first row.  
    \item After neglecting the higher-order term of the first row and multiplying each row by $[1/(2\tau)-1]$ as in Eq.~\eqref{eq:first order NS-part 2}, the first row becomes the viscous shear stress tensor $\sigma_{ij}$, having an order of magnitude denoted by $\mathcal{O}(\sigma_{ij})$. Note that $-\partial_{x_k} L_{ik}$ and $\rho a_i$ are the Lorentz force and body force, respectively, and have the same order of magnitude as $\mathcal{O}(\partial_{x_k} \sigma_{ik})$ according to force balance. Therefore, the order of magnitude of the second and third rows is $\mathcal{O}[(\tau-0.5)\Delta tu_j^{\rm eq}\partial_{x_k} \sigma_{ik}]$, which is smaller than $\mathcal{O}(\Delta x\partial_{x_k} \sigma_{ik})$ due to $\mathcal{O}(\tau-0.5)=\mathcal{O}(1)$ and $|u_j^{\rm eq}|<c$, and thus much smaller than $\mathcal{O}(\sigma_{ik})$ of the first row due to small $\Delta x$. Given the inevitable challenge in computing the second row about $L_{ij}$ , we can neglect the second and third rows, approximating $\sigma_{ij}$ by $[1/(2\tau)-1]\sum_\alpha e_{\alpha,i}e_{\alpha,j}f^{(1)}_\alpha$, and neglect $E_{ij}$ in computing $S_{ij}$ by Eq.~\eqref{eq:strain rate tensor for hydrodynamics}. 
    \item Although both the second and third rows are negligible compared to the first row, the third row can be exactly or partially cancelled out by other terms of Eq.~\eqref{eq:first order NS-part 2}. We therefore only neglect the second row about $L_{ij}$ and recover the momentum equation as accurately as possible.  
\end{itemize}

Now, combining the remaining terms of Eqs.~\eqref{eq:first order NS-part 1} and \eqref{eq:first order NS-part 2} for the momentum equation, we obtain: 
\begin{equation} \label{eq:first order NS}
\begin{split}
   \dfrac{\partial(\rho u_i^{\rm eq})}{\partial t}+\dfrac{\partial(\rho u_i^{\rm eq}u_j^{\rm eq})}{\partial x_j}=& -\dfrac{\partial p}{\partial x_i}-\dfrac{\partial L_{ij}}{\partial x_j}+\rho a_i+\dfrac{\partial}{\partial x_j}\left[\rho\nu(\dfrac{\partial u_j^{\rm eq}}{\partial x_i}+\dfrac{\partial u_i^{\rm eq}}{\partial x_j})\right] \\
   & +(\tau-\dfrac{1}{2})\dfrac{\partial}{\partial x_j}\left[\Delta t\rho(u_i^{\rm eq}a_j+u_j^{\rm eq}a_i)\right]-\tau\dfrac{\partial\Theta_{ij}}{\partial x_j} \\
   & -(1-\dfrac{1}{2\tau})\dfrac{\partial(\rho u_i-\rho u_i^{\rm eq})}{\partial t_0}-C\dfrac{\partial(\rho\Delta t a_i/2)}{\partial t_0}. 
\end{split}
\end{equation}
When the force model $F_\alpha^{\rm [e]}$ \cite{GuoPRE2002} with $C=[1-1/(2\tau)]$ and $\vec u^{\rm eq}=\vec u^{\rm F}=\vec v$ is applied, both the second and third rows of Eq.~\eqref{eq:first order NS} are zero and the reduced Eq.~\eqref{eq:first order NS} is the standard momentum equation. 

\section{Calculation of fluid-wall interaction force}\label{interaction force}
Due to the particle propagation feature of LBM, it is natural to consider the momentum exchange rate between fluid and wall surface \cite{WenJCP2014} in the interaction force calculation using the outgoing $\overline{f}_{\alpha'}$ and inwards bounced $\overline{f}_{\alpha}$. Considering a flat surface with $\vec n$ as the unit inward normal vector, a stress $P_i$ is defined here as the total momentum exchange per $\Delta t$ and per unit area $\Delta x^2$ and computed as: 
\begin{equation} \label{eq:momentum exchange}
\begin{split}
   P_i=\dfrac{\Delta x}{\Delta t}(\sum_{\vec e_{\alpha'}\cdot\vec n<0} e_{\alpha',i}\overline{f}_{\alpha'}-\sum_{\vec e_{\alpha}\cdot\vec n>0} e_{\alpha,i}\overline{f}_{\alpha}), \\
\end{split}
\end{equation}
where $\overline{f}_{\alpha'}$ and $\overline{f}_{\alpha}$ are the values recorded at the neighbouring fluid grid point $\vec x_{\rm f}$, namely $\overline{f}_{\alpha'}=f_{\alpha'}^{\rm rel}(\vec x_{\rm f},t)$ before starting propagation and $\overline{f}_{\alpha}=f_{\alpha}(\vec x_{\rm f},t+\Delta t)$ after a full-step propagation. When the wall is located a half-grid away from $\vec x_{\rm f}$, both values can be traced to the same wall surface location $\vec x_{\rm w}=\vec x_{\rm f}+0.5\Delta t\vec e_{\alpha'}$ at the same moment $\overline{t}=t+0.5\Delta t$, namely $\overline{f}_{\alpha'}=f_{\alpha'}(\vec x_{\rm w},t+0.5\Delta t)$ and $\overline{f}_{\alpha}=f_{\alpha}(\vec x_{\rm w},t+0.5\Delta t)$. For arbitrary wall-grid layouts or curved surfaces, the recorded distribution functions can be similarly traced forwards and backwards to the same imaginary point $\vec x_{\rm i}=\vec x_{\rm f}+0.5\Delta t\vec e_{\alpha'}$ at the same moment $\overline{t}=t+0.5\Delta t$, for which we have $\overline{f}_{\alpha'}=f_{\alpha'}(\vec x_{\rm i},t+0.5\Delta t)$ and $\overline{f}_{\alpha}=f_{\alpha}(\vec x_{\rm i},t+0.5\Delta t)$. In this case, $P_i$ computed using Eq.~\eqref{eq:momentum exchange} corresponds to the stress at $\vec x_{\rm i}$ rather than $\vec x_{\rm w}$, which should be consistent with a stress value extrapolated from the fluid side to $\vec x_{\rm i}$. Note that the recorded $\overline{f}_{\alpha'}$ and $\overline{f}_{\alpha}$ are different when the boundary is moving or not a half-grid away from $\vec x_{\rm f}$ \cite{LietalAMM2025}. 

Using the fact that $\vec e_{\alpha}\cdot\vec n$ only takes $-c$ and $c$ for the two groups with $\vec e_{\alpha}\cdot\vec n<0$ and $\vec e_{\alpha}\cdot\vec n>0$, respectively, Eq.~\eqref{eq:momentum exchange} can be reformulated as:   
\begin{equation} \label{eq:momentum exchange-r1}
\begin{split}
   P_i=-(\sum_{\alpha} e_{\alpha,i}e_{\alpha,j}\overline{f}_{\alpha})n_j, \\
\end{split}
\end{equation}
where the extra term $e_{\alpha,i}e_{\alpha,j}\overline{f}_{\alpha}n_j=0$ for $\vec e_{\alpha}\cdot\vec n=0$ is added to sum over all $\alpha$ for notation brevity. 

Similar to $\overline{f}_{\alpha}=f_{\alpha}(\vec x_{\rm w},t+0.5\Delta t)$ at $\vec x_{\rm w}$, we study the set of $\overline{f}_{\alpha}$ for any internal fluid grid point at $\vec x$. As previously used, the value sets before propagation and after a full-step propagation are $f_\alpha^{\rm rel}(\vec x,t)$ and $f_\alpha(\vec x,t+\Delta t)$, respectively. Then, the set of $\overline{f}_{\alpha}=f_\alpha(\vec x,t+0.5\Delta t)$ after a half-step propagation can be computed using a linear interpolation for $\vec x$: 
\begin{equation} \label{eq:f after half step}
\begin{split}
   \overline{f}_{\alpha}&=\dfrac{f_\alpha(\vec x+0.5\Delta t\vec e_\alpha,t+0.5\Delta t)+f_\alpha(\vec x-0.5\Delta t\vec e_\alpha,t+0.5\Delta t)}{2} \\
   &=\dfrac{f_\alpha^{\rm rel}(\vec x,t)+f_\alpha(\vec x,t+\Delta t)}{2}. \\
\end{split}
\end{equation}
We can define $\overline{\rho}=\sum_{\alpha}\overline{f}_{\alpha}$ and $\overline{u}_i=(1/\overline{\rho})\sum_{\alpha}e_{\alpha,i}\overline{f}_{\alpha}$, respectively, and correspondingly have $\overline{f}_{\alpha}^{\rm eq}=f_{\alpha}^{\rm eq}(\overline{\rho},\overline{u}_i)$ and $\overline{M}_{ij}^{\rm eq}$. Then, the stress tensor $-\overline{p}\delta_{ij}+\sigma_{ij}$ for hydrodynamic flows can be computed as $-\sum_{\alpha} (e_{\alpha,i}-\overline{u}_i)(e_{\alpha,j}-\overline{u}_j)\overline{f}_{\alpha}$ \cite{LiWang2010}, equivalently $\sigma_{ij}=-\sum_{\alpha} e_{\alpha,i}e_{\alpha,j}\overline{f}_{\alpha}^{(1)}$. The newly defined notations with an overline are introduced to emphasise the difference from their counterparts without an overline, except for $\sigma_{ij}$ that is the physical shear stress tensor but can be computed using either $f_{\alpha}$ or $\overline{f}_{\alpha}$. Generally speaking, we have:
\begin{equation} \label{eq:M of f after half step}
\begin{split}
   -\sum_{\alpha} e_{\alpha,i}e_{\alpha,j}\overline{f}_{\alpha}=-\sum_{\alpha} e_{\alpha,i}e_{\alpha,j}\overline{f}_{\alpha}^{(1)}-\overline{M}_{ij}^{\rm eq}=\sigma_{ij}-\overline{M}_{ij}^{\rm eq}. \\
\end{split}
\end{equation}
Applying Eq.~\eqref{eq:M of f after half step} to $\overline{f}_{\alpha}=f_{\alpha}(\vec x_{\rm w},t+0.5\Delta t)$, the stress $P_i$ computed using Eq.~\eqref{eq:momentum exchange} is a projection in the $\vec n$ direction, as shown in Eq.~\eqref{eq:momentum exchange-r1}, of a total stress tensor $\sigma_{ij}-\overline{M}_{ij}^{\rm eq}$ that includes four parts, namely the shear stress tensor $\sigma_{ij}$, the isotropic pressure tensor $-\overline{p}\delta_{ij}$, the momentum flux tensor $-\overline{u}_i\overline{u}_j\overline{\rho}$ if the boundary is moving, and the Maxwell stress tensor $-L_{ij}$ if $f_\alpha^{\rm eq}$ of the relaxation term is modified to include the Lorentz force for MHD flows. 

The outgoing $\overline{f}_{\alpha'}$ and inwards bounced $\overline{f}_{\alpha}$ used in computing $P_i$ are the values recorded at the neighbouring fluid grid point. Therefore, $P_i$ actually reflects the momentum exchange rate across a \textit{stationary} interface fixed to the grid point and contains $-\overline{u}_i\overline{u}_j\overline{\rho}$ when fluid moves together with the boundary. However, this momentum flux due to nonzero flow velocity is imperceptible to the wall surface since the wall and adjacent fluid are always moving at the same velocity. For hydrodynamic flows, the fluid-wall interaction force should only consist of $\sigma_{ij}$ and $-\overline{p}\delta_{ij}$, and including $-\overline{u}_i\overline{u}_j\overline{\rho}$ in the interaction force calculation violates Galilean invariance. Note that $-\overline{u}_i\overline{u}_j\overline{\rho}$ mainly affects the normal component of interaction force due to the non-slip velocity constraint and $\mathcal{O}(-\overline{u}_i\overline{u}_j\overline{\rho})\ll\mathcal{O}(-\overline{p}\delta_{ij})$ holds at low Mach numbers. A correction to $P_i$ should be made to offset $-\overline{u}_i\overline{u}_j\overline{\rho}$ for moving boundaries, as used in \cite{WenJCP2014}. For MHD flows, the additional $-L_{ij}$ contained in $P_i$ should also be cancelled out.  

Actually, Eq.~\eqref{eq:M of f after half step} also can be derived from the analysis of Section~\ref{MHD stress tensor}. Substituting Eq.~\eqref{eq:evolution falpha} to replace $f_\alpha^{\rm rel}(\vec x,t)$ and omitting the notation $(\vec x,t)$ as the default for brevity, Eq.~\eqref{eq:f after half step} can be rewritten into:
\begin{equation} \label{eq:f-neq after half step}
\begin{split}
   \overline{f}_{\alpha}-\overline{f}_\alpha^{\rm eq}&=\dfrac{f_\alpha^{\rm rel}-f_\alpha^{\rm eq}+f_\alpha(\vec x,t+\Delta t)-f_\alpha^{\rm eq}(\vec x,t+\Delta t)}{2} \\
   &=\dfrac{f_\alpha^{\rm rel}-f_\alpha^{\rm eq}+f_\alpha-f_\alpha^{\rm eq}}{2}+\mathcal{O}(\Delta t) \\
   &\approx(1-\dfrac{1}{2\tau})(f_{\alpha}-f_\alpha^{\rm eq})+\dfrac{F_\alpha}{2}, \\
\end{split}
\end{equation}
where $\overline{f}_\alpha^{\rm eq}$ is linearly interpolated for $t$ using $f_\alpha^{\rm eq}$ and $f_\alpha^{\rm eq}(\vec x,t+\Delta t)$, which are defined using $f_\alpha(\vec x,t+0.5\Delta t)$, $f_\alpha(\vec x,t)$ and $f_\alpha(\vec x,t+\Delta t)$, respectively. The second-order moment of Eq.~\eqref{eq:f-neq after half step} corresponds to: 
\begin{equation} \label{eq:second-order moment of f-neq after half step}
\begin{split}
   -\sum_{\alpha} e_{\alpha,i}e_{\alpha,j}\overline{f}_{\alpha}^{(1)}&=-(1-\dfrac{1}{2\tau})\sum_{\alpha} e_{\alpha,i}e_{\alpha,j}f_{\alpha}^{(1)}-\dfrac{\Theta_{ij}}{2} \\
   &=\sigma_{ij}-\dfrac{\Theta_{ij}}{2} \\
   &\approx\sigma_{ij},
\end{split}
\end{equation}
where the approximation for the third row is due to $\mathcal{O}(\Theta_{ij})\ll\mathcal{O}(\sigma_{ij})$, as discussed after Eq.~\eqref{eq:second-order moment of fneq}. Therefore, Eq.~\eqref{eq:second-order moment of f-neq after half step} obtained by a mathematical derivation following the Chapman-Enskog analysis is consistent with Eq.~\eqref{eq:M of f after half step} obtained by a physical reasoning of the momentum exchange process. 

It is noteworthy that $\overline{f}_{\alpha}=f_\alpha(\vec x,t+0.5\Delta t)$ of Eq.~\eqref{eq:f after half step} is a special quantity and different from $f_\alpha(\vec x,t)$ and $f_\alpha(\vec x,t+\Delta t)$ even at steady state. We denote the distribution functions after the relaxation and other local computations (i.e., before propagation) at two neighbouring grid points by $f_\alpha^{\rm rel}(\vec x,t)$ and $f_\alpha^{\rm rel}(\vec x-\Delta t\vec e_\alpha,t)$ that are different due to a spatial distance but unchanged after each full $\Delta t$ at steady state. At the grid point $\vec x$, the distribution function after a full-step propagation is $f_\alpha(\vec x,t+\Delta t)=f_\alpha^{\rm rel}(\vec x-\Delta t\vec e_\alpha,t)$. For an intermediate moment $t'\in(t, t+\Delta t]$, we apply a linear interpolation to obtain:
\begin{equation} \label{eq:f-intra during a timestep}
\begin{split}
   f_\alpha(\vec x,t')=f_\alpha^{\rm rel}(\vec x,t)+\dfrac{t'-t}{\Delta t}\left[f_\alpha^{\rm rel}(\vec x-\Delta t\vec e_\alpha,t)-f_\alpha^{\rm rel}(\vec x,t)\right], \\
\end{split}
\end{equation}
where $f_\alpha(\vec x,t')\to f_\alpha^{\rm rel}(\vec x,t)$ as $t'\to t^+$ is applied. On the other hand, a similar analysis for the last timestep indicates $f_\alpha(\vec x,t)=f_\alpha^{\rm rel}(\vec x-\Delta t\vec e_\alpha,t-\Delta t)=f_\alpha^{\rm rel}(\vec x-\Delta t\vec e_\alpha,t)$. Therefore, the \textit{underlying} time-evolution of the propagated $f_\alpha(\vec x,t')$ has a jump during each $\Delta t$, namely from $f_\alpha^{\rm rel}(\vec x-\Delta t\vec e_\alpha,t)$ at $t'=t$ to $f_\alpha^{\rm rel}(\vec x,t)$ at $t'\to t^+$. Physically speaking, the propagation process smoothly changes $f_\alpha(\vec x,t')$ from $f_\alpha^{\rm rel}(\vec x,t)$ to $f_\alpha^{\rm rel}(\vec x-\Delta t\vec e_\alpha,t)$ taking a timestep and then the relaxation process \textit{instantly} changes it back to $f_\alpha^{\rm rel}(\vec x,t)$. This intra-timestep discontinuity is invisible in usual LBM results that are computed only at the discrete moments with a full timestep. However, it leads to a correlation $-[1-1/(2\tau)]\sum_{\alpha} e_{\alpha,i}e_{\alpha,j} f_{\alpha}^{(1)}=\sigma_{ij}$ of Eq.~\eqref{eq:second-order moment of fneq}, which is different from $-\sum_{\alpha} e_{\alpha,i}e_{\alpha,j} \overline{f}_{\alpha}^{(1)}=\sigma_{ij}$ of Eq.~\eqref{eq:M of f after half step} that is consistent with the integral form of the continuous distribution function $f$.

\section{Total interaction force within a periodic unit}\label{total force}
In the calculation of total fluid-wall interaction force for a periodic geometry unit, we directly compute the sum of $P_i$ over the whole fluid-wall interface $\partial\Omega_{\rm fluid-wall}$ and denote the summation by a surface integral. Additionally, we denote the fluid areas at the periodic boundaries by $\partial\Omega_{\rm periodic-fluid}$ to obtain a closed surface integral over $\partial\Omega_{\rm fluid-wall}+\partial\Omega_{\rm periodic-fluid}=\partial\Omega_{\rm fluid}$, which is converted to a volume integral in the whole fluid domain $\Omega_{\rm fluid}$ with a negative sign due to using an inwards pointing $\vec n$ in Gauss's theorem:    
\begin{equation} \label{eq:total Pi}
\begin{split}
   \iint_{\partial\Omega_{\rm fluid-wall}}P_i \,{\rm d}A &= \oiint_{\partial\Omega_{\rm fluid}}P_i \,{\rm d}A - \iint_{\partial\Omega_{\rm periodic-fluid}}P_i \,{\rm d}A \\
   & =\iiint_{\Omega_{\rm fluid}}\dfrac{\partial}{\partial x_j}(\sum_{\alpha} e_{\alpha,i}e_{\alpha,j}\overline{f}_{\alpha}) \,{\rm d}V \\
   & \approx \iiint_{\Omega_{\rm fluid}}\dfrac{\partial}{\partial x_j}(M_{ij}^{\rm eq}-\sigma_{ij}) \,{\rm d}V \\
   & \approx \iiint_{\Omega_{\rm fluid}}\left[\rho a_i-\dfrac{\partial(\rho u_i^{\rm eq})}{\partial t}\right] \,{\rm d}V,
\end{split}
\end{equation}
where $\iint_{\partial\Omega_{\rm periodic-fluid}}P_i \,{\rm d}A=0$ is because $n_j$ of $P_i=-(\sum_{\alpha} e_{\alpha,i}e_{\alpha,j}\overline{f}_{\alpha})n_j$ takes opposite signs and meanwhile $-(\sum_{\alpha} e_{\alpha,i}e_{\alpha,j}\overline{f}_{\alpha})$ is unchanged at each pair of periodic boundaries, $\overline{M}_{ij}^{\rm eq}$ of the substituted Eq.~\eqref{eq:M of f after half step} is approximated by $M_{ij}^{\rm eq}$, and the small artefact terms of the recovered N-S equation~\eqref{eq:first order NS} are neglected. Therefore, the sum of $P_i$ for the total momentum exchange from fluid to wall is equal to the total external body force exerted on fluid at steady state. This is easy to understand from the LBM perspective that the momentum of fluid carried by $f_\alpha$ changes per timestep only through the bounceback process that results in $-P_i$ for each unit surface area and the force term that leads to $\rho a_i$ for each unit volume, having their sum equal to zero at steady state. However, further analysis is needed to understand the balance among various forces from the physical point of view.     

It is noteworthy that if the wall surface is not located a half-grid away from the neighbouring fluid grid points, Eq.~\eqref{eq:total Pi} still holds but the interface for the surface integral is located a half-grid away (see the discussion after Eq.~\eqref{eq:momentum exchange} using an imaginary point $\vec x_{\rm i}$) and the correspondingly enveloped domain for the volume integral is slightly different from the whole fluid domain without discretisation. Correspondingly, the sum of $P_i$ computed using the outgoing $\overline{f}_{\alpha'}$ and inwards bounced $\overline{f}_\alpha$ recorded at the neighbouring fluid grid points is equal to the total external body force exerted inside the enveloped domain at steady state. To be specific, there is a volume unit $\Delta x^3$ occupied by each fluid grid point and the enveloped domain is always the sum of volume units occupied by all fluid grid points. 

For hydrodynamic flows with static boundaries at steady state, we have $M_{ij}^{\rm eq}=p\delta_{ij}$ at boundaries and Eq.~\eqref{eq:total Pi} reduces to: 
\begin{equation} \label{eq:total hydrodynamic stress}
\begin{split}
    \iint_{\partial\Omega_{\rm fluid-wall}}(-p\delta_{ij}+\sigma_{ij})n_j \,{\rm d}A \approx \iiint_{\Omega_{\rm fluid}}\rho a_i \,{\rm d}V,
\end{split}
\end{equation}
which further clarifies that the total interaction force consists of the hydrodynamic pressure and viscous shear stress, and corresponds to the balance between the surface and body forces.  

For MHD flows, Eq.~\eqref{eq:total Pi} is still valid although $P_i$ additionally includes the Maxwell stress tensor for the Lorentz force. However, we might want to obtain the total interaction force due to the hydrodynamic pressure and viscous shear stress, which can be computed for static boundaries at steady state as follows: 
\begin{equation} \label{eq:total hydrodynamic stress for MHD}
\begin{split}
    \iint_{\partial\Omega_{\rm fluid-wall}}(-p\delta_{ij}+\sigma_{ij})n_j \,{\rm d}A \approx \iint_{\partial\Omega_{\rm fluid-wall}}L_{ij}n_j \,{\rm d}A +\iiint_{\Omega_{\rm fluid}}\rho a_i \,{\rm d}V,
\end{split}
\end{equation}
where the first term of the right-hand side (RHS) is the total Lorentz force exerted on fluid: 
\begin{equation} \label{eq:total Lij for MHD}
\begin{split}
    \iint_{\partial\Omega_{\rm fluid-wall}}L_{ij}n_j \,{\rm d}A=\oiint_{\partial\Omega_{\rm fluid}}L_{ij}n_j \,{\rm d}A=-\iiint_{\Omega_{\rm fluid}}\dfrac{\partial L_{ij}}{\partial x_j}\,{\rm d}V,
\end{split}
\end{equation}
where $\iint_{\partial\Omega_{\rm periodic-fluid}}L_{ij}n_j \,{\rm d}A=0$ is substituted due to $n_j$ taking opposite signs at each pair of periodic boundaries, and the negative sign is due to using an inwards pointing $\vec n$. Therefore, Eq.~\eqref{eq:total hydrodynamic stress for MHD} is the balance between the surface force and the two body forces. Note that the surface force for the fluid-wall interaction should consist of the hydrodynamic pressure and viscous shear stress while the sum of $P_i$ is the surface force minus the Lorentz body force, which is because the Lorentz force is embedded in the exchanged $\overline{f}_{\alpha'}$ and $\overline{f}_\alpha$ (thus in $P_i$) via $f_\alpha^{\rm eq}(\rho,\vec u^{\rm eq},\vec B)$ of the relaxation process.  

If both the Lorentz body force and external body force are implemented using a force model and therefore the equilibrium distribution function is unchanged, $-L_{ij}$ is not contained in $P_i$ and the first term of RHS of Eq.~\eqref{eq:total hydrodynamic stress for MHD} disappears. Nevertheless, the Lorentz force $-\partial L_{ij}/\partial x_j$ must be incorporated into $\rho a_i$ that becomes the total body force per unit volume. In this case, Eq.~\eqref{eq:total hydrodynamic stress for MHD} without the first term of RHS still represents the balance between the surface force and the two body forces.   

For MHD flows with insulating walls, the magnetic field at $\partial\Omega_{\rm fluid-wall}$ is constant $\vec B=\vec B_{\rm ext}$ and we denote $L_{ij}=(1/\mu)(B^2\delta_{ij}/2-B_iB_j)$ by $L_{ij}^{\rm ext}$. Therefore, the total Lorentz force of Eq.~\eqref{eq:total Lij for MHD} exerted on fluid is zero: 
\begin{equation} \label{eq:total Lij for MHD with insulating walls}
\begin{split}
    -\iiint_{\Omega_{\rm fluid}}\dfrac{\partial L_{ij}}{\partial x_j}\,{\rm d}V&=\iint_{\partial\Omega_{\rm fluid-wall}}L_{ij}^{\rm ext}n_j \,{\rm d}A \\
    &=\oiint_{\partial\Omega_{\rm fluid}}L_{ij}^{\rm ext}n_j \,{\rm d}A \\
    &=L_{ij}^{\rm ext}\oiint_{\partial\Omega_{\rm fluid}}n_j \,{\rm d}A \\
    &=0.
\end{split}
\end{equation}
Even though the total Lorentz force is zero, the local Lorentz force is nonzero and changes the local surface shear stress by force balance, which usually leads to an increase in the total surface shear stress for a given flow speed. 

For MHD flows with conducting walls, the total Lorentz force exerted on both fluid and solid phases within a periodic unit enveloped by the whole periodic boundaries $\partial\Omega_{\rm periodic}=\partial\Omega_{\rm periodic-fluid}+\partial\Omega_{\rm periodic-wall}$ is also zero: 
\begin{equation} \label{eq:total Lij for MHD in fluid and solid}
\begin{split}
    -\iiint_{\Omega_{\rm fluid}+\Omega_{\rm solid}}\dfrac{\partial L_{ij}}{\partial x_j}\,{\rm d}V=\oiint_{\partial\Omega_{\rm periodic}}L_{ij}n_j \,{\rm d}A=0, \\
\end{split}
\end{equation}
which is because $n_j$ takes opposite signs and meanwhile $L_{ij}$ is unchanged at each pair of periodic boundaries.  

While the computational domain is assumed as a fully periodic unit, the above analysis is readily generalisable to other cases where the domain is enclosed by a combination of periodic and insulating boundaries, or by fully insulating boundaries. For these scenarios, the total Lorentz force exerted inside the domain is always zero.

\section{Conclusions}\label{Conclusions}
 The shear stress tensor $\sigma_{ij}$ is studied using the Chapman-Enskog analysis of LBM for flows driven by external/internal body forces. Depending on the numerical scheme used to implement the body force, artefact terms might appear in the recovered N-S equations as well as in the calculation of $\sigma_{ij}$. Nevertheless, an order-of-magnitude analysis shows that the artefact terms are negligible in calculating $\sigma_{ij}$ for hydrodynamic and MHD flows. This clarifies the confusion in the literature, where both considering and not considering the artefact terms have been adopted without discussion.  
 
Different from the distribution function $f_\alpha$ after a full-step propagation, a special distribution function $\overline{f}_{\alpha}$ after a half-step propagation is defined and also can be used to calculate $\sigma_{ij}$. The same physical quantity $\sigma_{ij}$ can be computed using either $f_\alpha$ or $\overline{f}_{\alpha}$ with different coefficients, which is revealed by a further derivation based on the Chapman-Enskog analysis. The significance of introducing $\overline{f}_{\alpha}$ lies in that it explains why the fluid-wall interaction force can be computed using the intuitive momentum exchange rate during the halfway bounceback process, but not for other bounceback processes. This intuitively calculated interaction force also contains the momentum flux tensor for moving boundaries and the Maxwell stress tensor for some MHD flow simulations, both of which should be removed for an interaction force associated with the hydrodynamic pressure and viscous shear stress. 
 
%\section{References}

%\appendix{}
\end{document}